\documentclass[prl,aps,amsmath,superscriptaddress,showpacs,twocolumn]{revtex4}
\usepackage{epsfig}
\usepackage{float}
\usepackage{graphicx}
\begin{document}

\title{ Ab-initio calculations of Many-Body effects in liquids: \\
the electronic excitations of water}

\author{V. Garbuio}
\affiliation{European Theoretical Spectroscopy Facility (ETSF), and  CNR-INFM, Department of Physics,
 University of Rome
"Tor Vergata", Via della Ricerca Scientifica 1, 00133 Roma, Italy
}
\author{M. Cascella}
\affiliation{\'Ecole Polytechnique F\'ed\'erale de
Lausanne (EPFL), CH-1015 Lausanne, Switzerland}
\author{L. Reining}
\affiliation{European Theoretical Spectroscopy Facility (ETSF), and
Laboratoire des Solides Irradi\'es, \'Ecole Polytechnique, F-91128 Palaiseau, France}
\author{R. Del Sole} 
\affiliation{European Theoretical Spectroscopy Facility (ETSF), and  CNR-INFM, Department of Physics,
 University of Rome
"Tor Vergata", Via della Ricerca Scientifica 1, 00133 Roma, Italy
}
\author{O. Pulci}
\affiliation{European Theoretical Spectroscopy Facility (ETSF), and  CNR-INFM, Department of Physics,
 University of Rome
"Tor Vergata", Via della Ricerca Scientifica 1, 00133 Roma, Italy
}

\date{\today}
\begin{abstract}
\par
We present  {\it{ab-initio}} calculations of the excited state
properties of liquid water in the framework of Many-Body Green's
function formalism. Snapshots taken from
molecular dynamics simulations are used as input geometries to calculate electronic and optical spectra,
 and  the results are averaged
over the different configurations. 
% The electronic states are first
%obtained within the Density Functional Theory and then corrected
%within the so called "GW" approximation to take fully into account
%exchange and correlations effects.
 The optical absorption spectra with the inclusion of excitonic effects
are calculated by solving the Bethe-Salpeter equation. 
 These calculations are made possible by exploiting the
insensitivity of screening effects
 to a particular configuration. The resulting spectra are strongly modified by
many-body  effects,  both concerning peak energies and lineshapes, and are
in good agreement with experiments.
\end{abstract}

\maketitle The electronic structure of liquid water is still not fully
elucidated, even though it is essential to understand
the chemical and physical properties of many biochemical and
industrial processes that occur in solution, where it is crucial to
include  the role of the solvent in the reactions. Water is also
essential, both pure and as a solvent, for  living
organism survival and for biological systems in general. For these
reasons the study of the excited state properties of liquid water
is fundamental to advance in many research fields. However, in the
last years, the theoretical studies of liquid water \cite{Parrinello, Christiansen, Grossman, Sprik,
Sato, Silvestrelli, Chen, Raitieri, Berna, Galli}
have  mostly focused on its structure and ground state properties
whereas less effort has been dedicated to its electronic structure and optical absorption spectrum.
As a consequence,  experimental data about excited states are poorly understood.
One of the purposes of the present work
is to solve these issues  by carrying out    {\it{ab-initio}} many-body
calculations of the electronic structure and optical spectra of liquid water.

Here we generalize the application of the Many Body
Perturbation Theory (MBPT) 
\cite{Lucia} to liquid systems, and
present a calculation of the optical absorption
spectrum of liquid water from first principles,
 including both self-energy effects and the
electron-hole interaction. We show the occurrence of important excitonic
 effects, which are
 crucial for a good description and interpretation of experimental data.
\par\noindent
The main problem concerning the study of liquid water relies on
the fact  that, in order to simulate  a complex disordered system,
one should use a very large unit cell, with a consequent
prohibitive computational cost. In order to overcome this bottleneck,
 we 
used a smaller unit cell but exploiting several 
molecular dynamics (MD) snapshots of water as input geometries.
Averaging the resulting electronic and optical spectra over
many configurations should  give a good
approximation for the excited states properties of the real system.
We used 20 configurations of 17
water molecules in a cubic box with 15 a.u. side and 8 special
k-points. This is a quite low number of molecules
\cite{Galli}; still, the computational effort of the  many-body
 calculations for all the configurations would have been almost prohibitive. We will
show in the following that one can however restrict the calculation of certain ingredients, in particular of the screening, to a single configuration. In this way the calculations become feasible with a reasonable effort.
 Already the results obtained for 17
molecules, averaged
over several snapshots,  well compare with
%previous calculations carried out at the DFT level \cite{Galli} and with
experiments. \\

\begin{figure}[h]
\includegraphics[width=6cm, angle=0]{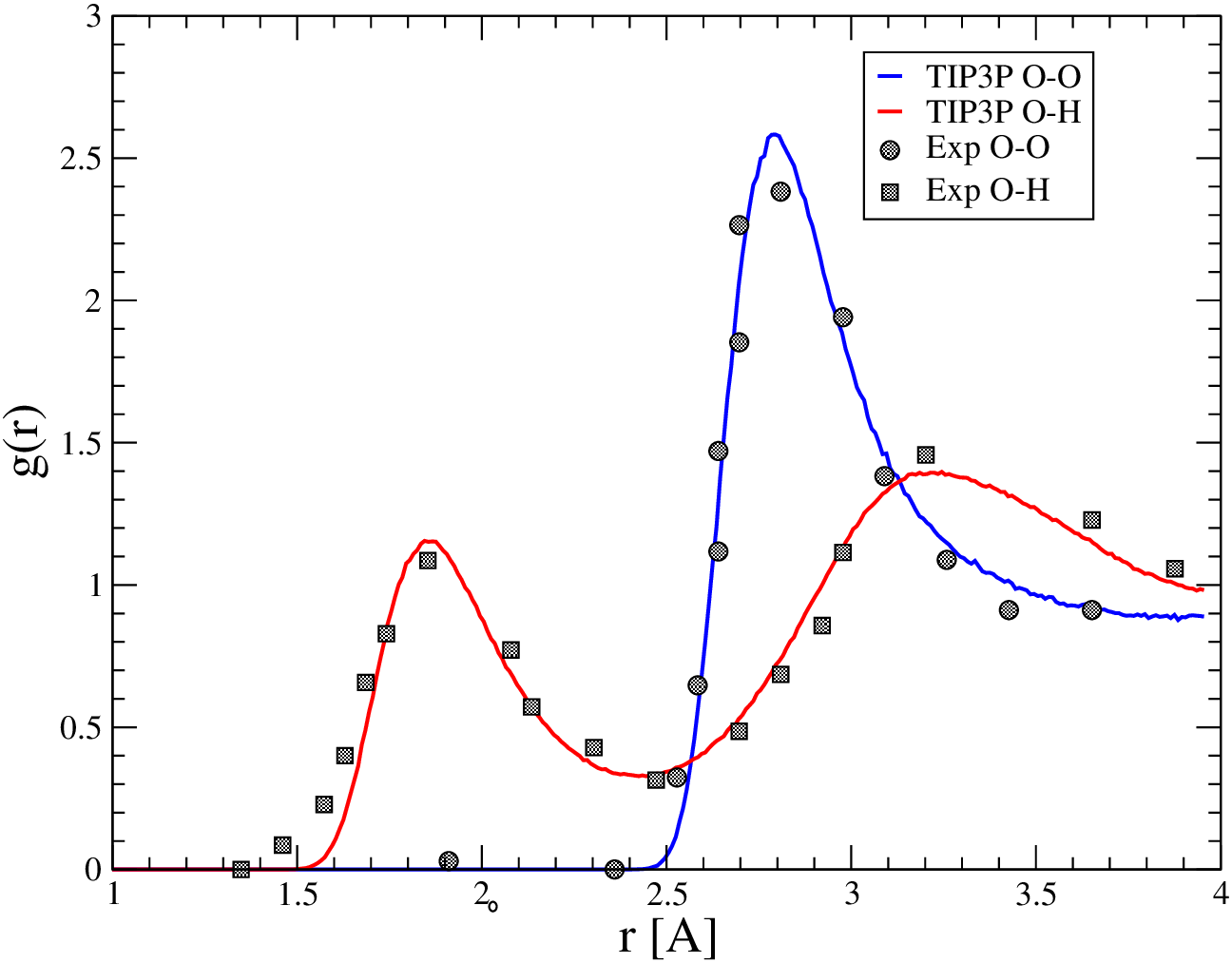}
\caption{\label{fig:gofr}
(Color online) Oxygen-oxygen (blue line) and oxygen-hydrogen (red line) radial distribution
functions averaged over the 20 molecular dynamics snapshots
compared to experimental data \cite{expgofr} (circles and squares).}
\end{figure}

The water configurations have been
obtained by sampling every 2 ns a 40 ns long classical MD simulation trajectory.
A TIP3P water model potential \cite{TIP3P} has been used to represent the water molecules in
our simulation box.
Equations of motion have been integrated numerically using a time-step of 1 fs.
The MD run has been done in the NVT ensemble, where
thermal equilibrium at 298 K has been achieved applying a Nos\'e-Hoover
thermostat\cite{nose}.
Despite the small size of our system, the resulting radial distribution
functions  are in very good agreement with the
experimental ones \cite{expgofr}, as it is shown in
Fig. \ref{fig:gofr}. This confirms that we are using good input
geometries for the excited state properties calculations.
\par
\par
We then performed the electronic and spectroscopic calculations on three levels: we started with density functional theory (DFT)
\cite{DFT-base},
\cite{LDA-GGA} to obtain the Kohn-Sham (KS) eigenvalues  and
eigenvectors. We hence constructed the DFT independent-particle absorption spectrum. As expected, it shows strong discrepancies with respect to experiment. These could not be cured by moving to Time-Dependent (TD) DFT, 
 in the adiabatic LDA (ALDA)\cite{Gross}. Therefore, we corrected the Kohn-Sham energy levels using the
Green's function perturbation approach, with the
exchange-correlation self-energy $\Sigma$ calculated within the GW
approximation $\Sigma=iGW$ \cite{Hedin},\cite{Lucia} ($G$ is
 the  one-particle Green's function and $W=\varepsilon^{-1}v$ the
screened Coulomb interaction). The Quasi-Particle (QP) energies
(that is, the
 electronic 'band structure' of water), were  calculated in first order perturbation theory:
\begin{equation}
\varepsilon^{QP}_n=\varepsilon^{DFT}_n+\Delta\varepsilon^{QP}_n=
\varepsilon^{DFT}_n +<n|\Sigma-V^{DFT}_{xc}|n>
\end{equation}
where $V^{DFT}_{xc}$ is the exchange and correlation KS potential
and  the self-energy is evaluated at the quasiparticle energy. As a last step,
 the optical
absorption spectrum was calculated
  by solving the Bethe-Salpeter equation
 \cite{Lucia}, so fully including local-field effects and the electron-hole interaction \cite{details}.

\begin{figure}[t]
\includegraphics[width=7cm, angle=0]{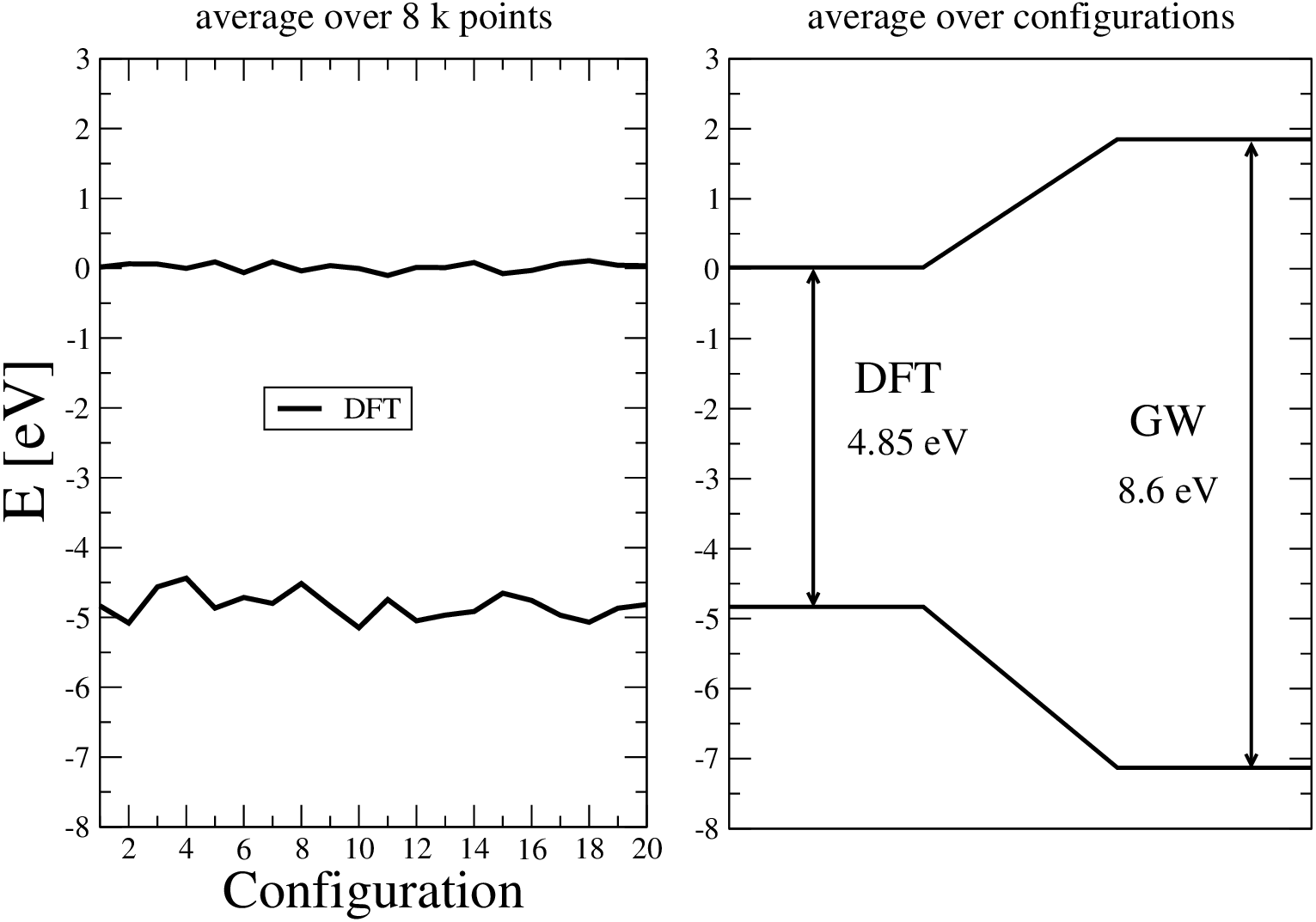}
\caption{\label{HOMO-LUMO}  Left panel: DFT HOMO and LUMO energies,
averaged over the 8 k-points, for
the 20 MD configurations. 
 Right panel: schematic  HOMO-LUMO gaps, calculated within DFT and GW,
 averaged over the 20 MD configurations.  }
\end{figure}

\begin{figure}[h]
\includegraphics[width=7cm, angle=0]{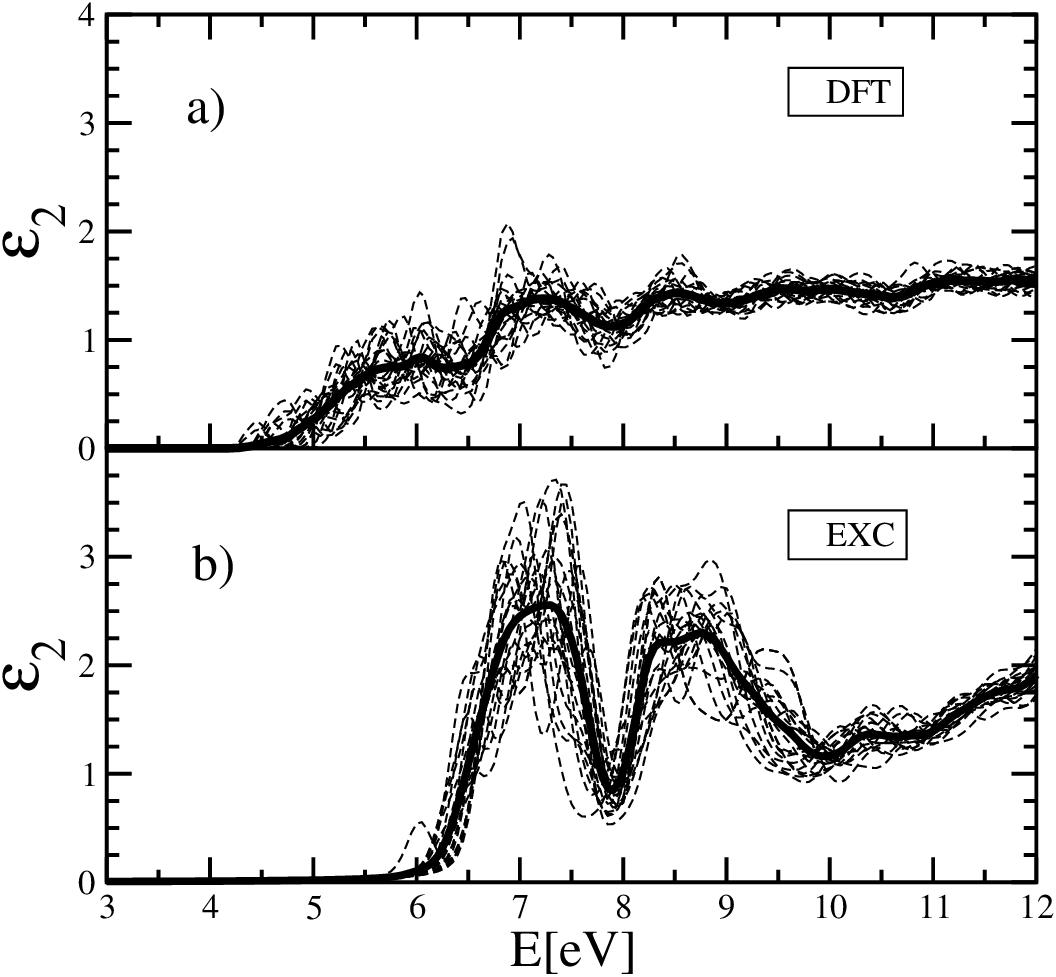}
\caption{\label{BSE} Optical absorption spectra of the 20
configurations (thin lines) and their average (thick lines). a): DFT single particle calculation; b):
 with the inclusion of self-energy  and electron-hole
interaction effects. }
\end{figure}

\begin{figure}[t]
\includegraphics[width=7cm, angle=0]{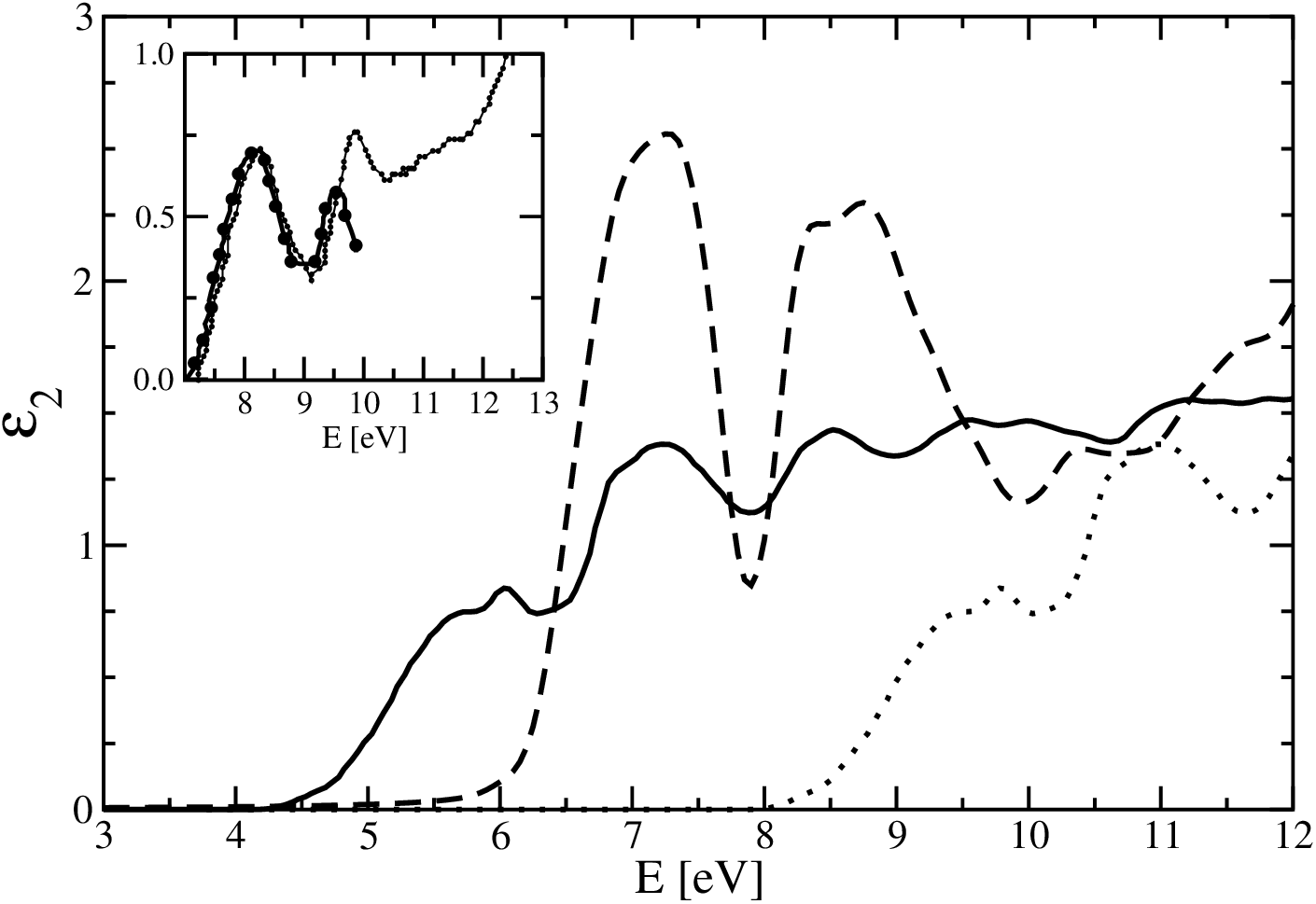}
\caption{\label{expabsspe} Calculated optical absorption spectra averaged
over 20 configurations. Solid line: single particle DFT spectrum; dotted line:
spectrum obtained with the inclusion of the GW corrections; dashed line:
spectrum obtained with the inclusion of the electron-hole interaction.
Inset: experimental optical absorption spectrum of
water (line: \cite{expABS}; circles: \cite{expABS2}). }
\vskip 1.0cm
\end{figure}

DFT-KS results for the electronic and spectroscopic properties of
water (obtained averaging over the 8 k points for each of the 20 configurations) are shown in
Figs. \ref{HOMO-LUMO} and  \ref{BSE}a.  The configuration-averaged 
HOMO-LUMO gap
turned out to be 4.85 eV, in good agreement with previous DFT
calculations  \cite{Parrinello} but strongly
underestimating the experimental gap (8.7 $\pm 0.5$  eV
\cite{expgap}), as expected in DFT calculations.
Also optical absorption spectra at the DFT independent-particle level
 obtained for the 20 MD snapshots and, more important, their average
 (Fig. \ref{BSE}) do not compare in a
satisfactory way with the experimental absorption spectrum
\cite{expABS}, \cite{expABS2}  reported in the inset of Fig.\ref{expabsspe}:
we can observe that  the onset of the absorption is
strongly underestimated in our calculation, the peak positions are red-shifted
in comparison to the experiment, and the relative intensities
of the first two absorption peaks are not well reproduced.

\begin{figure}[h]
\includegraphics[width=7cm, angle=0]{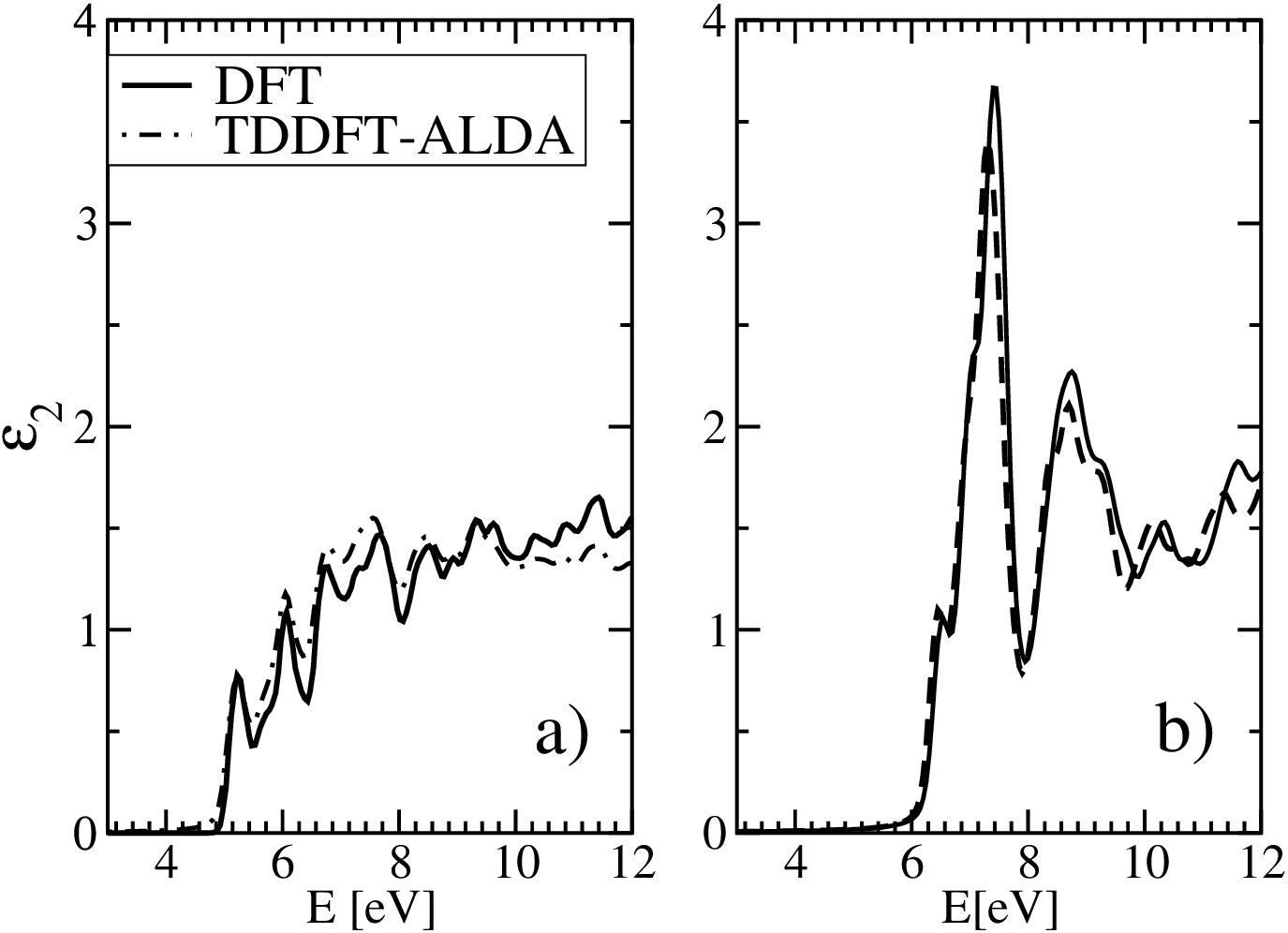}
\caption{ \label{TDLDA} Optical absorption spectrum for one MD configuration.
a) Time Dependent DFT within the Adiabatic LDA vs single particle 
Kohn-Sham  DFT spectrum. b) BSE spectrum
 calculated using the full screened electron-hole interaction (solid line)
 and using for W a constant dielectric
function $\varepsilon^{-1}=1/1.7$ (dashed line).}
\end{figure}

We have therefore performed calculations of the optical absorption spectrum  within
TDDFT \cite{Gross}. In principle this
represents an exact  way to calculate optical spectra, but the quality of the results depends on the approximation that is used to describe exchange-correlation effects. A widely used and computationally most efficient approximation is 
ALDA. Unfortunately our results, shown in Fig. \ref{TDLDA}a, shows 
 no improvement with respect to the static DFT independent-particle result. Long-range and/or dynamical effects that are missing in the ALDA kernel should hence be important for the absorption spectrum of water, and one has to resort to 
more elaborate (hence time-consuming) TDDFT approximations \cite{kernelTDDFT},
 or to work in a different framework. We have adopted the latter choice in this work,
 since Many-Body Perturbation Theory offers a well established 
 way to compute and analyze optical absorption spectra of extended systems. 

 In order to correct the KS electronic gap and optical spectra, one can calculate the
GW corrections $\Delta \epsilon^{QP}_n$ to the KS energies. This should be done 
for $all$ the 20 MD configuration, followed by an average. 
In particular the calculation of the screened Coulomb interaction for 20 
configurations constitutes however a true bottleneck. Instead, one can imagine that 
a change in configuration does not modify drastically 
 the  screening  
 of an additional hole or electron. In fact, changing for example the position 
in space of an occupied orbital also moves the region of strong screening in 
the same direction. One can therefore hope that GW corrections $\Delta \epsilon^{QP}_n$ are
 quite stable with respect to the configuration. 

This actually turned out to be true in our calculations. In other words, although the configurations
were different, and so were the KS eigenvalues, the difference between DFT and GW, $\Delta
GW=\varepsilon_n^{QP}-\varepsilon_n^{DFT}$,
 was practically constant going
from one snapshot to another. This is shown in table
\ref{table:gap} for three configurations. We could, hence, use the
same GW corrections for all the DFT configurations.
\par
\begin{table}[h]
\begin{tabular}{|c|c|c|c|c|}\hline
 & DFTgap& $\Delta GW$ HOMO  & $\Delta GW$ LUMO  &  $\Delta GW$ gap  \\ \hline
 E19 & 4.38 & -2.02 & 1.83 &  3.85 \\ \hline
 E01  & 4.72 & -2.03 & 1.82 &  3.85\\ \hline
 E13  & 4.21& -2.04 & 1.82 &  3.86\\ \hline
\end{tabular}
\caption{\label{table:gap} Test
GW corrections to the HOMO and LUMO energy levels and
to the electronic gap, for  three different water configurations (E19, E01, E13) \cite{note_tab}. Energies  in eV.
}
\vskip 1.0cm
\end{table}
\par

 The GW corrections increase the electronic HOMO-LUMO
gap to 8.6 eV (see Fig. \ref{HOMO-LUMO}) \cite{GWdetails}, well within the
experimental range \cite{expgap}. (Note that a larger gap,
$\sim$ 9.5 eV, is obtained in hexagonal ice \cite{Hahn}. The disordered band-edge landscape leads indeed to a shrinkage of the minimum gap with
respect to the reference ordered system, ice in this case.). Since
the calculated GW shift is almost constant for all the bands, GW
optical spectra show lineshapes very similar to DFT ones,
 but shifted to higher energies, as shown in Fig. \ref{expabsspe}. The
agreement with the experimental $\varepsilon_2(\omega)$ is hence not at all
improved, since the relative intensities of the two
absorption structures (seen in experiments at 8.3 and 9.6 eV) are
still not reproduced, and their position, from being  red shifted
in DFT calculations (Fig.\ref{expabsspe}, solid line), are now strongly blue
shifted in GW (Fig.\ref{expabsspe}, dotted line).
% A comparison of the GW spectrum
%to that of hexagonal ice \cite{Hahn}, shows a different lineshape,
% such as the first GW peak occurs 9.5 eV, with an onset at 8.6 eV, while in ice the first peak
%is at about 13 eV, with and onset at 9.5 eV. When we say that GW corrections are
% smaller in liquid water than in hexagonal ice, refer to the different positions
% (9.5 eV versus 13 eV) of the first peaks in the single-quasiparticle absorption spectrum.

We have hence clear hints
that single particle and single quasi-particle approaches are not
sufficient to describe the optical properties of water, and that
it is necessary to include  the
electron-hole interaction. To this end we had 
 to solve the Bethe
Salpeter equation, where electrons and holes interact  through the screened Coulomb potential {\it W}. These cumbersome calculations had to be  done
for
all the 20 MD configurations, in order to obtain an average.
%\vskip 0.5cm

Our results are
shown in Fig. \ref{BSE} for all the 20 MD configurations and in Fig. \ref{expabsspe},
 where  their average is reported (dashed line). Dramatic many-body effects are evident. 
Agreement with
 experiment both in  energy peak positions and onset, as well as in
the relative intensities of the first two peaks, is significantly improved. 
The main remaining discrepancy is an overall redshift, that might be  
due to the fact that our GW calculations are not self-consistent but use 
DFT wavefunctions and energies.  The first peak in the spectrum turns out to be a
 bound exciton with a
binding energy of 2.4 eV and large oscillator strength.  These are
a consequence of the weak electronic dielectric screening of water
($\varepsilon_{\infty}\sim 1.8$).  The second peak results from an
excitonic enhancement of the  oscillator strength of interband transitions
with respect to the single
quasi-particle case.
The binding energy of the lower exciton, although quite large, is
 smaller than
 the value $E_b$= 3.2 eV  found in
hexagonal ice \cite{Hahn}. One can in fact imagine that the mixing of electron-hole pairs of different energy, which leads to the bound exciton, partially counterbalances the disorder effect on the quasiparticle gap \cite{Elliot} (namely the gap shrinking that is determined by a local increase of occupied state  energies, and a local decrease of unoccupied state energies).
%The lower intermolecular distances in the liquid phase with respect to ice
%(which are responsible of the puzzling behaviour of the liquid being heavier
%than the solid) allow the electron and the hole of the exciton to be
%more delocalized, and lower its binding energy.
Moreover, the higher density in the liquid phase with respect to
ice (around 7$\%$), 
 may also  play a  role in reducing the
exciton binding energy, by allowing a greater mobility of
electrons and holes.

It is of fundamental importance to explore also  to which extent a
 detailed calculation of each single configuration is mandatory. In fact, 
the effect of the electron-hole interaction does not 
depend significantly on details of the screened Coulomb 
interaction  $W=\varepsilon^{-1}v$ \cite{Lucia}, but rather on macroscopic averages. 
Therefore, and as we have explicitly verified, the optical
spectra of the various MD configurations can be calculated using
for the electron-hole interaction the same screening,
$\varepsilon^{-1}(q,\omega)_{G,G\prime}$, obtained for any of the
independent MD snapshots. Moreover, test calculations done using
a {\it constant} value for $\varepsilon^{-1}$ have shown that a
good agreement can also be obtained using a value for
$\varepsilon^{-1}=1/1.7$, (very near to the experimental macroscopic electronic value 1/1.8) as shown in Fig.\ref{TDLDA}b.

Our findings concerning the stability of GW and BSE calculations  may have important consequences on
future calculations for water (and maybe  for other liquids), since the ab-initio evaluation of
$\varepsilon^{-1}(q,\omega)_{G,G\prime}$ is the real bottleneck of
many-body calculations, and its evaluation for many different
snapshots makes this approach too cumbersome for becoming a state
of the art method. Hence,
 to be able
to determine accurate optical spectra for many configurations,
using the same dielectric constant
to screen the electron-hole interaction,
 will enormously speed up the calculations.

 In conclusion, we have presented an {\it ab-initio} many-body
calculation of the electronic structure and optical properties of 
liquid water obtained averaging the results of
snapshots taken from molecular dynamics simulation.
% We used a
%relatively small cell since calculations based on Green's function
%perturbation theory have a almost prohibitive computational cost.
%In order to try to reproduce the 'real' disordered system, we have
%calculated electronic and optical properties for 20 snapshots
%taken from molecular dynamics, and then averaged.
We have found that the GW corrections are almost independent of
the particular snapshot considered and give an average HOMO-LUMO electronic
gap of 8.6 eV,  smaller than that of hexagonal ice \cite{Hahn},
and that the effect of the screened electron-hole interaction is
the same for all the configurations. This result implies a huge
reduction of the computational effort, and opens a pathway to low
cost calculations on other disordered systems.

Absorption spectra calculated with the inclusion of excitonic
effects show important structures related to the electron-hole
interaction which are essential to obtain a good description of
experimental data. The onset position and the relative intensities
of the first two peaks are well reproduced {\it only if} the
electron-hole interaction is fully taken into account. ALDA calculations
do not improve at all the DFT spectra.

This work has been supported by the EU's 6th Framework Programme
through the NANOQUANTA NoE
(NMP4-CT-2004-500198). Bethe-Salpeter calculations have been performed using the EXC code http://www.bethe-salpeter.org/.
 We acknowledge MIUR for financial support (NANOSIM and PRIN2005), and INFM for CINECA cpu time.
We thank P. Carloni, Ari P. Seitsonen and M. Marsili for useful discussions.


\begin{thebibliography}{100}
\bibitem{LDA-GGA}  We perform pseudopotential DFT simulations both in the GGA-PW91 and LDA approximations; although the resulting
gaps show differences of about 0.25 eV, the final GW results  are
independent of the starting point, as expected.
 We sample
 the Brillouin zone with a 8 k points mesh and   use a kinetic energy cut-off of 50 Ry.
\bibitem{details}
 Optical absorption spectra at the BSE level have been obtained
 including all the filled states and 100 empty levels. The resulting 20 excitonic Hamiltonians 
(one each MD configuration)
 sized 54400x54400,
have been diagonalized using the Haydock algorithm.
\bibitem{note_tab} The planewave cutoff energy has been lowered by 40\% for these tests. The $\Delta GW$ corrections 
in table \ref{table:gap}
results  to be   0.1 eV larger than the well converged values obtained  with the higher cutoff.
\bibitem{GWdetails} GW corrections have been calculated using 19933 plane waves for the exchange part of $\Sigma$, and 13997 plane waves
 for its correlation part. The screened $W$ has been calculated within the plasmon pole model, using  600 empty bands.
\end{thebibliography}
\end{document}